\documentclass[review]{elsarticle}

\usepackage{geometry}
\usepackage{amsmath,amsfonts}
\usepackage{amssymb}
\usepackage{txfonts}
\usepackage{subfigure}
\usepackage{booktabs}
\usepackage{multirow}
\usepackage[table]{xcolor}
\usepackage{array}
\usepackage[caption=false,font=normalsize,labelfont=sf,textfont=sf]{subfig}
\usepackage{verbatim}
\usepackage{graphicx}

\usepackage{newtxtext,newtxmath} 
\usepackage{booktabs}

\begin{document}

\begin{frontmatter}


\title{Research progress on intelligent optimization techniques for energy-efficient design of ship hull forms}


\author[inst1]{Shuwei Zhu$^{1}$, Siying Lv$^{1}$, Kaifeng Chen$^{1}$, Wei Fang$^{1,}$*, Leilei Cao$^{2,}$}

\address{%
$^{1}$ \quad School of Artificial Intelligence and Computer Science, Jiangnan University, Wuxi, 214122, China\\
$^{2}$ \quad Innovation Center of Yangtze River Delta, Zhejiang University, Jiashan 314100, China\\
$^{3}$ \quad School of Intelligent Manufacturing, Nanjing University of Science and Technology, Nanjing, 210094, China}

\begin{abstract}
The design optimization of ship hull form based on hydrodynamics theory and simulation-based design (SBD) technologies generally considers ship performance and energy efficiency performance as the design objective, which plays an important role in smart design and manufacturing of green ship. An optimal design of sustainable energy system requires multidisciplinary tools to build ships with the least resistance and energy consumption. Through a systematic approach, this paper presents the research progress of energy-efficient design of ship hull forms based on intelligent optimization techniques. We discuss different methods involved in the optimization procedure, especially the latest developments of intelligent optimization algorithms and surrogate models. Moreover, current development trends and technical challenges of multidisciplinary design optimization and surrogate-assisted evolutionary algorithms for ship design are further analyzed. We explore the gaps and potential future directions, so as to paving the way towards the design of the next generation of more energy-efficient ship hull form.
\end{abstract}


\end{frontmatter}


\section{Introduction}

Shipping exhibits a great impact on the global emissions, which accounts for almost 3$\%$ of the global anthropogenic emissions in 2018, and leads to the increase of greenhouse gases (GHG) emissions by about 9.6$\%$ degree from international shipping compared to the 2012 levels \cite{trivyza2022decision}. Recently, the International Maritime Organisation (IMO) made a target to reduce 50$\%$ GHG emissions by 2050. The challenging target has attracted increasing attention towards ship energy systems and alternative fuels that can offer a more sustainable performance over the past decade.
The optimization of an energy system either on land or on ships can be considered at three levels--i.e., synthesis, design, and operation \cite{frangopoulos2020developments}, and each level cannot be completely isolated from the others. To be more specific in shipping areas, it is suggested in \cite{frangopoulos2020developments, esmailian2019novel} that, the optimization of the whole ship can be considered as a complex system with all its subsystems--e.g., hull form, energy equipments, propulsors, navigation equipment, etc., which is indispensable for its whole life cycle. Thus, it is necessary and beneficial to further expend the border of ship energy system to include aspects belonging to other disciplines, such as hull form and propulsor optimization, and hence, moving towards the holistic ship optimization \cite{frangopoulos2020developments}.

Nowadays, smart design of ship hull form has attracted much attention under the promotion of low-carbon economic. The hull form design does not focus merely on static water resistance minimization any longer, but shifts towards pursuing optimal overall navigation performance. Under the concept of green ship design, it is of great demand to establish the energy-efficient and environmental-friendly ship manufacturing industry. In 2014, the IMO proposed the Energy Efficiency Design Index (EEDI), which calls for hull form design to be safer, more environmental-friendly, more cost-effective, and more comfortable in the future. Therefore, energy conservation and emission reduction have become the main issue of future development of ship design. As the core component of ship design, hull form design plays a crucial role in energy conservation and fuel efficiency, which is considered as the fundamental technology that can bring profound and long-lasting impacts in terms of energy efficiency \cite{tadros2023review}. Hence, energy-efficient hull form design plays an important role to reduce total resistance, fuel consumption, and carbon dioxide emission. However, traditional hull form design methods struggle to meet the development demands of green ship.

Ship hull form design calls for multidisciplinary optimization tools involving numerous sub-domains, with close relation to hydrodynamic performance, navigation performance, energy efficiency, seakeeping
performance, general layout, operational economy, etc. With the development of computer aided design techniques, the simulation-based design (SBD) approaches have significantly promoted the transformation of hull form design from traditional empirical methods to intelligent modes \cite{chi2016overview}. Through the SBD approaches, ship performance calculation and optimization techniques based on computational fluid dynamics (CFD) \cite{diez2018stochastic, zhang2018hull, lin2019automatic,miao2020cfd, nazemian2022multi} are combined with hull geometry reconstruction methods, aiming to achieve the best performance (e.g., minimal resistance and lowest energy consumption) under given conditions or constraints. 

This paper reviews the progress of intelligent optimization techniques for ship hull form design in recent years. Specifically, we elaborate the latest development of intelligent optimization methods and surrogate-assisted optimization in hull form design. Moreover, some insights for exploring the new generation of intelligent hull form design optimization are provided.

\section{Main technologies in hull form optimization}

Ship hull form design optimization based on hydrodynamics theory and SBD technology is a complicated system engineering, which integrates multi-disciplines, i.e., CFD, CAD, computer technologies, and optimization methods. General hull form design patterns are shown in Figure 1. The basic principle of ship hull form design based on SBD technology is to complete the numerical simulation and hydrodynamic calculation for given performance targets such as ship resistance and sea-keeping with the assistance of CFD technology \cite{islam2022comparison}. Then, optimization methods are used to search the hull geometric design space based on the automatic hull geometry reconstruction technology so as to obtain the excellent hull form with the best hydrodynamic performance under condition of comprehensive constraints. Therefore, hull form optimization system mainly involves five key technologies: CFD numerical simulation technology, hull geometry reconstruction technology, optimization technology, approximation technology and integration technology.

\begin{figure}[h]
\begin{center}
\includegraphics[scale=0.55]{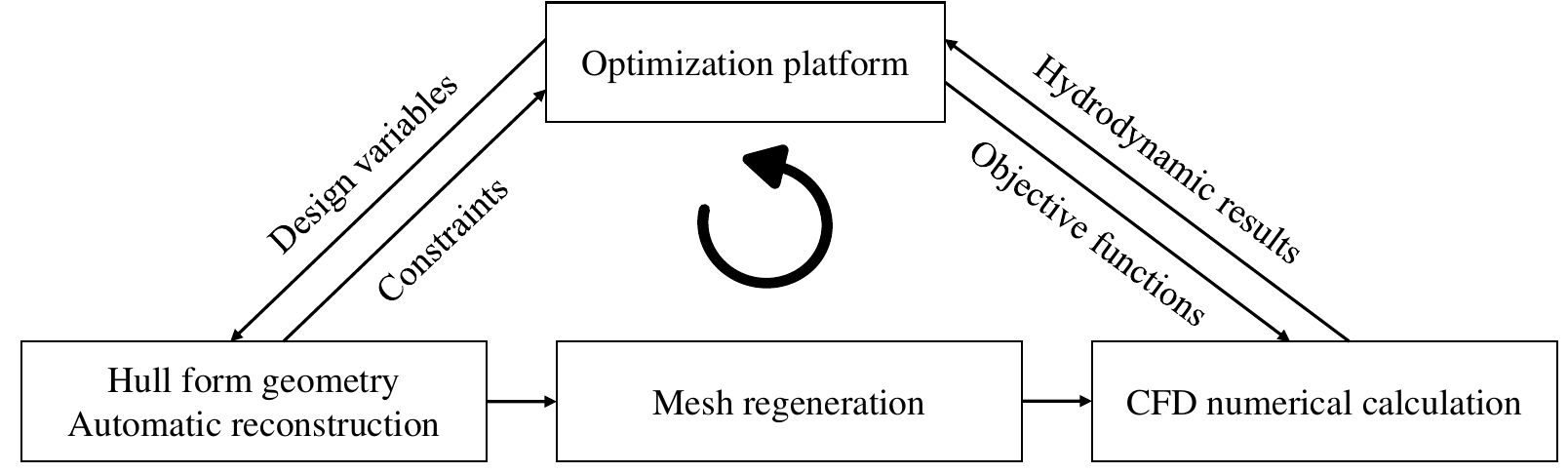}
\caption{Principle of ship hull form design optimization based on SBD technology.}
\label{Fig1}
\end{center}
\end{figure}

\subsection{CFD numerical simulation technology}

Ship hull form optimization based on hydrodynamic theory requires detailed description of the flow field around the hull and effective measurements to control the flow around. Therefore, ship hull form optimization is inseparable from the scientific guidance of ship hydrodynamic theory, including the calculation method and analysis technology of hydrodynamics. The hydrodynamic theory used to solve ship resistance can be divided into two categories, i.e., potential flow theory and viscous flow theory, while potential flow theory can be further separated into linear one and nonlinear one. Currently, commonly-used potential flow theories include linear wave resistance theory based on Michell integral method and nonlinear wave resistance theory based on Rankine source method. Viscous flow theory primarily utilizes CFD methods to predict a ship's viscous resistance and wave resistance. It involves setting up numerical simulations for static water conditions and wave conditions to investigate the optimal hull design with minimal resistance.

\subsection{Hull geometry reconstruction technology}

Hull geometry reconstruction technology serves as a bridge between ship resistance performance evaluations and optimization methods. For hull form optimization based on hydrodynamic theory, especially CFD methods \cite{islam2022comparison}, the relationships between objective functions (such as commonly-used minimum total resistance) and design variables are often implicit. How to establish the relationships between design variables and objective functions is a prerequisite for implementing CFD-based hull form optimization. Usually, a small number of parameters are used to parameterize the geometric shape of the hull form firstly. Then, we build the relationships between hull's shape parameters and design variables. After that, better design variables can be obtained with the assistance of optimization methods. Lastly, geometric reconstruction techniques can be used to modify the geometric shape of the ship hull. According to different ship parameters, hull geometric reconstruction techniques can be divided into two categories: (1) ship parameterization--i.e., expressing the geometric shape of the hull through a series of ship characteristic parameters; (2) geometric modeling--i.e., achieving the reconstruction of the hull's geometric shape by changing the positions of a series of control points. It is worthy noting that the stability of the designed hull form needs to be guaranteed when generating different geometric shapes.

\subsection{Optimization technology}

Generally, the optimization techniques commonly used in the field of ship design can be broadly classified into three categories: 
\begin{enumerate}
\item  Gradient-based optimization algorithms \cite{zhang2019research}, including Nonlinear Programming (NLP), Sequential Quadratic Programming (SQP), and Mixed Integer Quadratic Programming (MISQP).
\item Derivative-free intelligent optimization algorithms \cite{larson2019derivative, serani2016ship}, such as Genetic Algorithm (GA) \cite{zhao2020heuristic,zhao2021iterated,esmailian2022ship}, Particle Swarm Optimization (PSO) algorithm \cite{serani2016ship, hu2022conceptual}, Artificial Bee Colony (ABC) algorithm \cite{zhu2019optimized} and other evolutionary algorithms.
\item Mixed optimization algorithms combining gradient-based optimization with intelligent optimization techniques. e.g. combining GA with SQP \cite{qian2012ship}, combining DPSO with SQP \cite{kim2016hull}, and combining DPSO with DIRECT method \cite{serani2016ship,hou2017minimum}. 
\end{enumerate}

Gradient-based optimization methods show fast convergence and good performance given initial values, but there also has obvious shortcomings. Ship hull form optimization usually involves multiple disciplines such as speediness, wind and wave resistance, maneuverability, etc. It is difficult to establish clear analytical expressions between each performance index and design variables while they may possess the characteristics of multi-peak, non-conductive or black box. The gradient information can only be obtained by computationally expensive numerical analysis. For complex and nonlinear ship resistance performance optimization problems, gradient-based optimization cannot be applied directly and it is easy to converge to the local optimal solutions. Moreover, optimization results obtained by it are very sensitive to the settings of initial values. Contrarily, intelligent optimization algorithms have strong global search ability, but their local search ability is poor, which results in slow search speed. Therefore, the combination of above two optimization methods can take advantage of their respective advantages to form a more efficient hybrid global optimization algorithm.

\subsection{Approximation technology}

Optimization algorithms usually require numerous iterative objective evaluations before achieving satisfactory results. It intends to be unaffordable if a high-precision solver is involved, e.g., CFD, which requires long response time and intensive computational cost. Therefore, it is difficult to complete a quick optimization process within the stipulated time. How to solve a large number of numerical calculations based on hydrodynamic theory is a prerequisite for ship hull form optimization engineering. There are two main methods: (1) high-performance computing--e.g., parallel/distributed computing techniques; (2) approximation technology, it can simulate the design space to obtain the implicit expressions of multiple objective functions with low computational cost, and thus greatly reducing the computational resource in the optimization process. As a result, approximation technique has received more attention from researchers. Approximation technique mainly includes some parts \cite{viana2021surrogate}: (1) Screening and variable reduction, (2) design of experiments (DoE) sampling, (3) approximate model (or surrogate model) construction, and (4) sequential sampling. 
\begin{enumerate}
\item Screening and variable reduction is an efficient step for reducing the cost of the surrogate’s construction, with drastic dimensionality reductions being possible.

\item DoE is a crucial part in building surrogate models, which is an effective mathematical statistical sampling method for selecting sample points for simulation. It is important to allocate sample points in the design space reasonably so as to improve the model accuracy. A good experimental design should satisfy that the sample points are full of the design space, that is, uniformity and orthogonality, and can effectively compress the sample size to reduce computational overhead. 

\item At present, commonly-used surrogate models include Kriging model, Response Surface Method (RSM), Neural Network (NN), Radical Basis Function (RBF), and Support Vector Machine (SVM).  Compared with other commonly-used surrogate models, Kriging models can provide not only the predicted value of the prediction point, but also the prediction error (variance), that is, the confidence interval of the Gaussian process (GP) regression model. Thereby, the prediction probability errors given by Kriging model can be used naturally to dynamically update the model in order to improve its prediction accuracy.

\item Sequential sampling is an efficient way of making use of limited computational budget. Techniques make use of both the prediction and the uncertainty estimates of the surrogate models to intelligently sample the design space.

\end{enumerate}

\subsection{Integration technology}

Hull form optimization based on hydrodynamic theory is a complex system engineering involving various technologies. How to integrate different modules to form a unified interface optimization platform is also the key part of realizing optimization process automation. Nowadays, integration technologies are mainly completed based on optimization platforms, such as ISIGHT \cite{tezdogan2018investigation} and OPENFOAM \cite{he2019design}. ISIGHT is a popular and relatively mature optimization platform, and thus most researchers use it for synthesis \cite{tezdogan2018investigation, zhang2018computational, zong2018hull, lin2019automatic, guan2021parametric, hu2022conceptual, luo2021hull, yang2020improved}. ISIGHT optimization platform integrates CFD resistance calculation module and CAD geometry reconstruction module, and reserves program interface between modules. OPENFOAM is a CFD open source platform that can be re-developed, which has been used a lot \cite{huang2016hull, coppede2019hydrodynamic, rafiee2022multi, liu2023multi, villa2021effective, ccelik2021reduced}. Wan \textit{et al}. \cite{wu2017neumann} developed the naoe-FOAM-SJTU solver for marine ship hydrodynamics problems based on the OPENFOAM platform. Furthermore, they have developed a holistic hull optimization platform for hull form design, called OPTShip-SJTU \cite{miao2020cfd, liu2021hull, liu2022resistance}, which integrates several key components like hull form expression, ship shape transformation, experimental design, hydrodynamic performance evaluation, approximation module and optimization algorithms.

\section{Application of SBD and intelligent optimization methods for energy-efficient ship hulls}

This section reviews and analyzes the research on ship hull form design based on SBD technology and intelligent optimization methods, elaborates and summarizes the main theories and methods involved in optimization, and focuses on the latest developments of intelligent optimization methods and surrogate model design in this field. In general, ship hull form optimization design problem can be expressed as follows:
\begin{equation}
    \begin{aligned}
        \mathrm{Min} : &f({\bf{x}}),  x\in D\\
        \mathrm{s.t.} : &g_i({\bf{x}})\leq 0, {i}=1,\dots,{p}\\
                         &h_j({\bf{x}})\leq 0, {j}=1,\dots,{q}\\
    \end{aligned}
\end{equation}
where $f(\textbf{x})$ is the objective function of hull form optimization, i.e., hydrodynamic performance index, such as resistance value, companion fraction value, and motion response. And $\textbf{x}$ refers to design variables (solution vectors), representing the ship shape transformation parameters, $D$ is the feasible domain, which means the design space of the problem. Any point on the feasible domain represents a solution vector $\textbf{x}$. $g_i(\textbf{x})$ and $h_j(\textbf{x})$ are inequality constraint function and the equation constraint function \cite{peng2022constrained}, respectively, e.g., precise shape constraints for the wet surface area, main scale, displacement volume and even propeller shaft height. Therefore, ship hull form optimization is a typical class of engineering optimization problems that involves a large number of design variables and constraints.

Ship hull form design based on SBD solved by intelligent optimization algorithms requires a large number of computationally expensive objective functions before finding the global optimum. It is extraordinarily time-consuming if physical model tests or numerical simulation calculations are carried out in each generation of the optimization process. Therefore, numerous surrogate models, instead of the original high-fidelity CFD evaluations, can be used to reduce computational cost. However, surrogate model accuracy plays a crucial role in affecting the effectiveness of the optimization. Often, surrogate model reliability can be improved through cross-validation and reduction of model variance. DoE generally makes the sample points fill the design space, which does not consider the differences of the objective function in the design space. As a result, it cannot guarantee the local accuracy of surrogate models. Dynamically updating the approximation model by adding new sample points in an orderly manner with certain criteria (e.g., selecting points with largest variance) can effectively improve the surrogate accuracy. Therefore, the optimization idea of ship resistance performance of SBD technology combining intelligent optimization algorithms and approximate models is shown in Figure 2, where approximate models replace part of true objective function calculations, and the optimal hull form design can be obtained with minimal computational resource.

\begin{figure}[h]
\begin{center}
\includegraphics[scale=1.3]{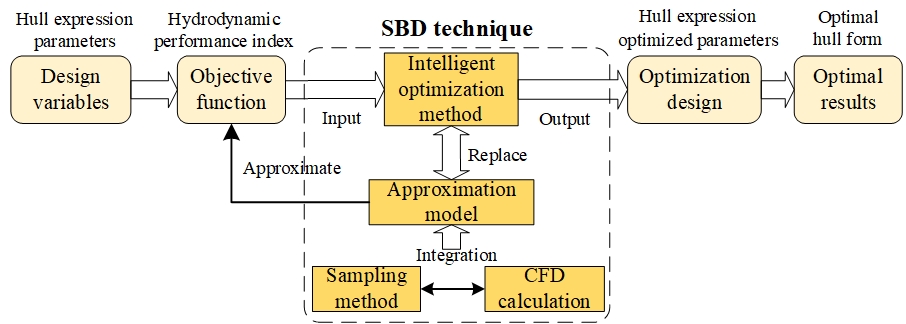}
\caption{The framework of SBD technology incorporating intelligent optimization and approximate models.}
\label{Fig2}
\end{center}
\end{figure}

We review the representative publications on optimal design of ship hull form published in important scientific journals, which are indexed by Web of Science in the past ten years. As can be seen from Figure 3, publication count keeps on the rising trend over time, especially in the last three years. Note that, the publications of 2023 is collected until September, but it has the potential to exceed the count of the years before. Therefore, we summarize the mainstream methods proposed in the representative publications, as discussed according to the major components depicted in Figure 2.

\begin{figure}[h]
\begin{center}
\includegraphics[scale=0.65]{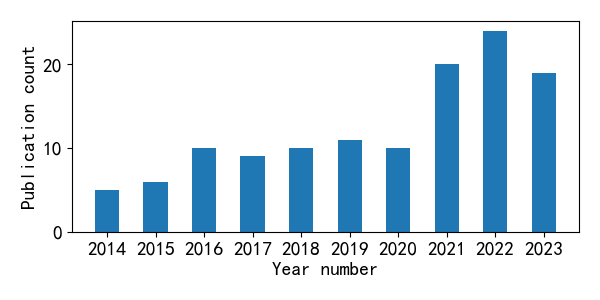}
\caption{Distribution of articles published in important journals.}
\label{Fig3}
\end{center}
\end{figure}

\subsection{Design variables based on hull form expressions and transformations}

As the first step of hull form optimization design, the key technology of hull form expression lies in how to design variables effectively, accurately and directly to express the complex hull geometry (e.g., ship scale ratio, longitudinal centre of buoyancy, and etc.). Based on ship form representation technology, ship form transformation technology mainly focuses on which deformation method (hull geometry reconstruction technology) can effectively obtain smooth and practical ship form, large deformation space and fewer deformation control parameters. The choice of hull form transformation technique and the determination of deformation parameters affect the design space for hull form optimization design problems. The current ship hull form transformation methods include \cite{wang2022improved} translation method, free-form deformation (FFD), B-splines, Lackenby transform, superposition fusion (morphing), RBF, orthogonal basis functions and so on. FFD method has been used to automatically modify the geometry of the ship \cite{villa2021effective, miao2020cfd, liu2022resistance,yang2018integrated, ni2020multiple, li2022application, hamed2022multi, wang2021many}. In the literature \cite{coppede2019hydrodynamic}, multiple branches of ship surfaces are combined and deformed in conjunction with the FFD method. In the literature \cite{zha2021hull}, RBF method was used for local deformation of the bow, and FFD method for local deformation of the stern, and translation method for global deformation, respectively. 

Meanwhile, some recent research focuses on the establishment of ship hull form transformation methods, such as \cite{villa2021effective, wang2022improved, li2022application, cheng2018hull, ichinose2022method}. Among them, Li \textit{et al}. \cite{li2022application} combined the mesh deformation with the adaptive method. Wang \textit{et al}. \cite{wang2022improved} proposed an improved RBF method for ship hull form expression and optimization. Ichinose \cite{ichinose2022method} expanded the traditional method which only superimposes two or three hull forms to $N$ hull forms by introducing the center of gravity coordinate system in the superposition fusion method, and further assisted decision making by visualizing and analyzing the decision space in the vicinity of the optimal hull forms.

In order to obtain a more practical and efficient design space for hull form representation, some dimension reduction techniques have been used to simplify the design variables \cite{ccelik2021reduced, d2020design, zheng2021dynamic, khan2022shape, liu2021linear, qiang2022optimization} and reduce computational cost to some extent. Some works use Karhunen-Loeve Expansion (KLE) \cite{d2020design, diez2015design, serani2022hull, chen2015high, diez2014uncertainty,chang2023research} or Proper Orthogonal Decomposition (POD) \cite{liu2023multi, villa2021effective} to compress dimensionality of design space, and then establish a reduced dimensional representation for hull form transformation. Among them, D'Agostino \textit{et al}. \cite{d2020design} proposed offline decision space dimension approximation method that relies on geometric variance to conduct decision space evaluations without simulation or performance analysis. It is worthy pointing out that, under certain conditions, KLE (or POD) method is approximately equivalent to principal components analysis (PCA) method \cite{diez2014uncertainty}. Zheng \textit{et al}. \cite{zheng2021dynamic} proposed a dynamic spatial approximation method that used data mining techniques to dynamically approximate the range and number of design variables during the optimization process. Qiang \textit{et al}. \cite{qiang2022optimization} designed a multi-stage spatial approximation technique by combining Self-Organizing Map (SOM) and rough set theory respectively. Zheng \textit{et al}. \cite{zheng2021dynamic} established a hierarchical dimension approximation method by combining SOM with simulated annealing search method. Khan \textit{et al}. \cite{khan2022shape} proposed a ship hull shape-supervised initial design space dimensionality approximation method to describe the important intrinsic structure of the hull by constructing shape-signature vectors (SSVs) such that the resulting reduced dimensional subspace retains the required reconfiguration capabilities, which can provide diversity, robustness, and more important hull physicsal information.

\subsection{Objective functions based on hydrodynamic performance evaluation}

The technology of ship hydrodynamic performance evaluation is to provide calculation tools for the objective functions (such as speediness, seakeeping and maneuverability) in ship form optimization design problems, so as to evaluate the quality of the optimization results. The reliability, efficiency, and economy of hydrodynamic performance evaluation techniques are major concerns and difficulties for ship hull form optimization design. Traditional methods use empirical formulas and model tests, while some recent methods adopt CFD numerical simulation calculations. They can be  divided into following two categories:

\begin{enumerate}
\item Methods based on potential flow theory. Michell integral method, Neumann-Kelvin method, Dawson method, and Rankine method are used to calculate the resistance of a rising wave. Slice theory and facet method are adopted for calculating seakeeping performance. Also, slender body theory and facet method can be used to calculate maneuvering performance. At present, there have been many studies on the optimal design of ship hull form based on the potential flow theory. Moreover, the calculation cost is low as a result of the small mess volume. However, the optimized ship shape may be more complicated and strange as a result of the ignorance of viscous effects. In addition, practical problems are simplified and resulting calculation accuracy may not meet the actual requirements, which may account for the failure of its reliability.

\item Methods based on viscous flow theory. For example, Reynolds Average Navier-Stokes (RANS) equation considers viscosity, separated eddy simulation method, direct numerical simulation method. Ship hull form optimization viscous solvers generally use RANS models to meet the high accuracy requirements. The numerical results obtained by viscous flow CFD methods are usually of high accuracy and it can capture many details of the flow field.
\end{enumerate}

\subsection{Intelligent optimization algorithm}

Early ship hull form optimization concentrates on single-objective optimization \cite{wang2023study}, and its aim is to minimize the hydrodynamic resistance performance. 

Recently, ship hull form design optimization turns to multi-objective optimization which considers several performance indicators simultaneously \cite{wang2023study_2}. For example, Yang \textit{et al}. \cite{yang2018integrated} improved the hydrodynamic performance of a ship hull form by solving the hull design optimization problem at different speeds, and explores the design space using a multi-objective PSO (MOPSO) optimization algorithm where two objective functions - i.e., total drag at two different speeds (12 and 14 knots) - are evaluated by RANS solver. Following the same idea, Ni \textit{et al}. \cite{ni2020multiple} used MOPSO algorithm to optimize the SWATH hull form where total drag at two different speeds (11 and 15 knots) are considered.

Compared with single-objective optimization problems, the main characteristic of multi-objective optimization Problems (MOPs) is that their optimization objectives are conflicting \cite{zhou2018kernel,zhu2021new}. That is to say, there is no single optimal solution to make all objectives optimal, and the optimal solution is a set of solutions that compromise each other among objectives, i.e., Pareto front (PF). The aim of solving a MOP is to obtain a set of vectors that are uniformly distributed and as close as possible to the true PF of the problem, which in turn facilitates the decision maker's selection based on preferences or further requirements \cite{zhu2018many,zhu2020evolutionary,zhu2021hierarchical}. Therefore, it is required that the resulting final solution set has good convergence and diversity. As a class of global optimization algorithms based on population-directed stochastic search, Multi-objective Evolutionary Algorithms (MOEAs) have become the most popular methods for solving MOPs. We list some representative MOEAs, such as NSGA-II based on Pareto dominance \cite{deb2002fast}, MOEA/D based on decomposition strategy \cite{zhang2007moea} and the multi-objective particle swarm optimization algorithm \cite{coello2004handling}.

During the past decade, ship researchers have carried out a lot of research on ship optimization design using SBD technology. The various intelligent optimization algorithms used for ship hull form design in the existing mainstream literature are summarized in Table 1. Also, the comparative distribution of the number of different algorithms (the improved algorithms for PSO, IPSO, and DPSO are also denoted by PSO) is shown in Figure 4. It can be seen that the single-objective optimization methods, such as PSO and its improved versions IPSO and DPSO, and the multi-objective optimization methods, such as NSGA-II, have received more attention. In addition to consider two objective functions in ship hull form optimization design, some recent works explore three or even more objective functions, and then use reference point-based high-dimensional multi-objective optimization algorithm NSGA-III to solve them \cite{deb2013evolutionary}.

\begin{table}[ht]
\small
\begin{center}
\caption{Intelligence optimization algorithms used in the literature of hull form optimization.}
\begin{tabular}{ccc} \hline
Objective num & Algorithms & Articles \\  \hline
single-objective & GA & \cite{liu2022multi,esmailian2022ship,wang2023study}\\
& MIGA & \cite{zong2018hull, lin2019automatic, wen2022optimal} \\
& MPGA & \cite{yang2022new} \\
& ABC & \cite{zhu2019optimized, huang2016new} \\
& PSO &  \cite{serani2016ship, hu2022conceptual, qiang2022optimization,chen2015high,wang2021aerodynamic,zhang2022research,wu2021uncertain,diez2010robust,wei2019hull}\\
& DPSO & \cite{kim2016hull, hou2017minimum,serani2016ship,d2020design, serani2016parameter} \\ \hline
multi-objective & MOGA  & \cite{larson2019derivative,cheng2019multi,skoupas2019parametric}\\
& NSGA-II  & \cite{wu2017neumann, miao2020cfd,liu2022resistance, wang2022improved,zha2021hull,cheng2018hull,qiang2022optimization,wan2022interval,miao2020hull,guan2022hull,guo2020cfd,jung2019hull,feng2018multidisciplinary,mao2022multi,liu2017multiple,lin2018hull,mittendorf2021hydrodynamic,tian2021multi,alam2015design} \\
& NSGA-III & \cite{wang2021many,hamed2022multi,nazemian2022multi}\\
& MOPSO  & \cite{yang2018integrated,ni2020multiple,diez2010robust,diez2018stochastic}  \\
& MODPSO & \cite{serani2022hull,pellegrini2017formulation} \\
& MOEA/D & \cite{zakerdoost2019multi} \\
& MOABC  & \cite{huang2016hull,jiang2023multi} \\
& DMOEOA &  \cite{chen2021multi}\\
& SHERPA & \cite{nazemian2021cfd} \\
\hline
\end{tabular}
\end{center}
\end{table}

\begin{figure}[h]
\begin{center}
\includegraphics[scale=0.6]{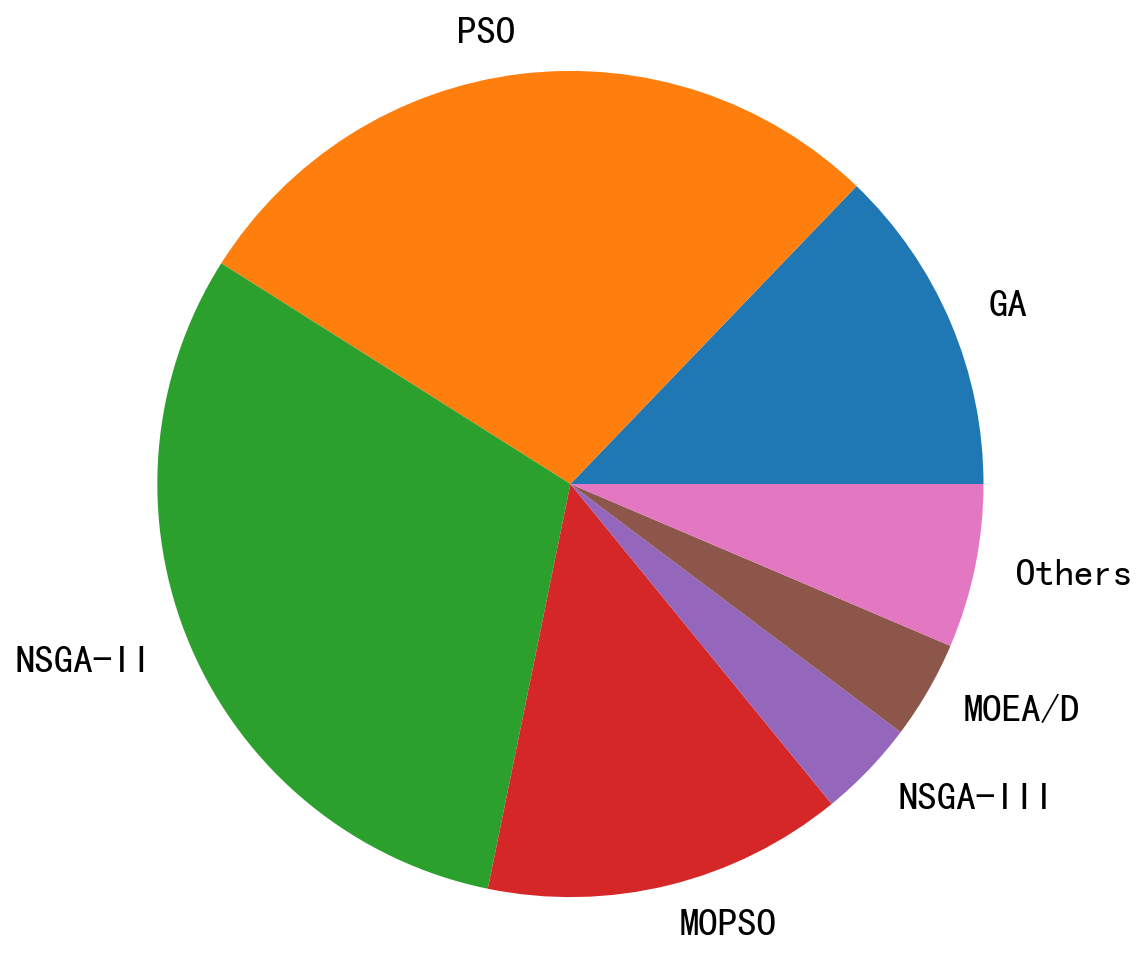}
\caption{Distribution of number of different intelligent optimization algorithms for ship hull-form design.}
\label{Fig4}
\end{center}
\end{figure}

\subsection{Approximate technology}

The accuracy of any approximate model is primarily affected by two factors: (i) noise in the data and (ii) inadequacy of the fitting model (called modeling error or bias error). In view of the first factor, screening and variable reduction as well as DoE techniques can play important roles in constructing a good approximation model.

As the number of variables in the surrogate increases, the number of simulations required for surrogate construction rises exponentially (curse of dimensionality). However, some of the variables may have only a negligible effect on the response surface approximation. Hence, several techniques have thus been proposed for evaluating the importance of the variables economically \cite{viana2021surrogate}, like variable screening techniques, variance-based techniques, variable transformation techniques and dimensionality reduction by subspace construction.

Mainstream DoE methods include orthogonal experimental sampling \cite{tahara2008computational,tahara2011single}, Latin Hypercube Sampling (LHS) \cite{diez2014uncertainty,feng2018multidisciplinary,mao2022multi} and Optimized Latin Hypercube Sampling ( Optimized Latin Hypercube Sampling, OLHS) \cite{lin2019automatic,liu2023multi,zhang2018hull} and Sobol sampling \cite{chang2016sample,gaggero2022marine}. For example, Tahara \textit{et al}. used an orthogonal test sampling method to sample the design space for the optimization of DTMB 5415 \cite{tahara2008computational} and Delft ship hull form \cite{tahara2011single}, respectively. Huang \textit{et al}. \cite{huang2016hull} used the LHS method for sampling the central slice optimization of a trimaran. Wu \textit{et al}. \cite{wu2017neumann} used OLHS sampling for optimizing design of DTMB 5415. In addition, some other sampling methods have been successfully applied in ship hull form optimization. Volpi \textit{et al}. \cite{volpi2015development} used the Latin center-of-mass Voronoi gridded sampling method to initially sample the Delft ship hull form optimization. Ouyang et al \textit{et al}. \cite{ouyang2023application} proposed an improved maximum entropy sampling method for S60 ship hull form for optimization analysis. We summarize DoE methods used in the existing literature in Table 2. It can be clearly seen that LHS and OLHS methods are most popular among researchers because they have better prediction accuracy and simpler construction process. Moreover, it is evident that Sobol sampling attracts several attention mainly due to the fact that Sobol sampling can effectively ensure the construction of high-dimensional surrogate models with high prediction accuracy.

\begin{table}[ht]
\small
\begin{center}
\caption{Design of experiment approaches used in the literature of hull form optimization.}
\begin{tabular}{cc} \hline
DoE sampling method & Articles \\  \hline
Orthogonal test sampling & \cite{tahara2008computational,tahara2011single} \\
LHS &  \cite{tezdogan2018investigation,huang2016hull,coppede2019hydrodynamic,ccelik2021reduced,diez2014uncertainty,wu2021uncertain,wan2022interval,feng2018multidisciplinary,mao2022multi,mittendorf2021hydrodynamic,chang2016sample,volpi2015development,ye2020optimal,renaud2022multi}\\
OLHS & \cite{zhang2018computational,lin2019automatic,luo2021hull,wu2017neumann,liu2021hull,liu2022resistance,miao2020hull,lin2018hull,tian2021multi,zakerdoost2019multi,chen2021multi,zhang2018hull} \\
Sobol sampling & \cite{guan2021parametric,villa2021effective,ccelik2021reduced,liu2022multi,wang2021aerodynamic,chang2016sample,gaggero2022marine,furcas2020design} \\
Uniform design sampling & \cite{wang2021many,wang2023study} \\
Sequence sampling based on Voronoi diagrams & \cite{yang2020improved} \\
Latin Center of Mass Voronoi Gridded Sampling & \cite{volpi2015development} \\
Maximum Entropy Sampling  & \cite{ouyang2023application} \\
\hline
\end{tabular}
\end{center}
\end{table}

Recently, surrogate model construction in hull form optimization has sprung up a large number of research works. Peri \textit{et al}. \cite{peri2001design} introduced surrogate models into the field of hull form optimization earlier, and compared the advantages and disadvantages of various surrogate models such as RSM, Kriging, NN and RBF. Since then, several studies have given examples of hull form optimization based on surrogate models. Tahara \textit{et al}. \cite{tahara2008computational} used Kriging model in optimizing the Delft catamaran. Volpi \textit{et al}. \cite{volpi2015development} used DRBF and DKG methods to construct approximation models to optimize the Delft catamaran. Chen \textit{et al}. \cite{chen2015high} adopted four approximation models, i.e., RBF, Kriging, SVM, and multi-harmonic spline methods, and experimental results showed that RBF model works better. Leotardi \textit{et al}. \cite{leotardi2016variable} used RBF to construct an approximate model for optimizing DTMB 5415. Diez \textit{et al}. \cite{diez2015multi} adopted NN method to train an approximate model in optimizing DTMB 5415. Serani \textit{et al}. \cite{serani2016ship} used RBF method to construct a surrogate model to optimize DTMB 5415. Huang \textit{et al}. \cite{huang2016hull} used RBF method to construct an approximate model in the triple-hulled ship optimization. Li \textit{et al}. \cite{li2014bow} used RSM in optimizing the total resistance of a bulk carrier. Chen \textit{et al}. \cite{chen2015high} used RBF to optimize the Delft Catamaran 372 model. Yang \textit{et al}. \cite{chi2016overview} used RBF in optimizing a Series 60 hull to reduce the total drag at two speeds. Wu \textit{et al}. \cite{wu2017neumann} took advantage of Kriging method to construct an approximate model to optimize the DTMB 5415. Hou \cite{hou2017minimum} used NN to optimize the EEDI of a Wigley hull. Diez \textit{et al}. \cite{diez2018stochastic} used RBF to optimize the drag and seakeeping performance of Delft Catamaran 372 model considering stochastic conditions. Zong \textit{et al}. \cite{zong2018hull} used second order RSM to optimize the total drag coefficient of a trimaran. Coppede \textit{et al}. \cite{coppede2019hydrodynamic} used Kriging model to optimize the total drag of KCS vessel. Zhang \textit{et al}. \cite{zhang2018computational} used Elman neural network to optimize the total drag coefficient of DTMB-5512 and Wigley III hulls in calm water at given speed. Miao \textit{et al}. \cite{miao2020cfd} optimized the S60 catamaran using the Kriging model to reduce drag by varying the de-hulling shape and separation. Serani \textit{et al}. \cite{serani2022hull} optimized the DTMB-5415 hull by using RBF to minimize the expected value of the mean total drag and maximize the maneuverability of the vessel in a completely stochastic environment. The popularity of deep learning in recent years also brings opportunities in the field of ship optimization, e.g., Zhang \textit{et al}. \cite{zhang2022research} constructed a surrogate model for predicting the total resistance based on a deep belief network and showed its superiority over the traditional surrogate models.

The methods for evaluating hydrodynamic performance of ship have different fidelity levels. It is possible to obtain hydrodynamic performance metrics for new hull forms by using different physical models or numerical discretization schemes if these methods complement each other. Then, high-precision surrogate models can be constructed and thus evaluating new sample hulls in a highly efficient manner. Therefore, for simulation-based hull hydrodynamic performance optimization, the establishment of multi-fidelity surrogate models is very necessary. A series of research works have followed this line of idea and have become one of the research hotspots \cite{liu2023multi,liu2022multi,alam2015design,zakerdoost2019multi,gaggero2022marine,pellegrini2022multi,piazzola2023comparing,pellegrini2023multi}. For example, a dual-fidelity Co-Kriging surrogate model was proposed for the optimization design of marine propellers, where the boundary element method is used for the low-fidelity model and RANS is used for the high-fidelity model \cite{gaggero2022marine}. Liu \textit{et al}. \cite{liu2022multi} proposed a multi-fidelity Co-Kriging surrogate model that uses more low-fidelity sample data to assist fewer high-fidelity sample data to predict high-fidelity outputs, which can reduce the total computational cost and make the surrogate model have relatively high accuracy. A multi-fidelity Co-Kriging surrogate model was similarly developed for Japanese bulk carriers in \cite{liu2023multi}, and POD technology is used to study the flow field dimensionality reduction so as to make full use of the flow field results of the new sample hull. In conclusion, the multi-fidelity surrogate models can balance high efficiency and high accuracy while using fewer samples for high-fidelity simulations and more samples for low-fidelity simulations. However, it has not yet been widely applied in the optimization of hydrodynamic performance of ship hull forms.

The accuracy and applicability of surrogate models may vary for different engineering problems. Sometimes, high-fidelity accuracy surrogate cannot be guaranteed with only a single model in the entire design space. Therefore, surrogate ensemble (SEN) methods have been proposed to fully utilize advantages of different surrogate models. To be specific, ensemble surrogates are constructed by weighted combination of several different surrogate models \cite{goel2007ensemble}. The key point of constructing an ensemble model is to determine the weight coefficients. In \cite{hu2022conceptual}, an adaptive surrogate ensemble strategy that combines Polynomial Response Surface (PRS) and Kriging surrogate models by means of weighted sums was proposed. The weight of each surrogate model is determined according to its prediction error. The optimal weighting factor in surrogate ensemble is determined based on the minimization of the local mean square error in \cite{ye2020optimal}. Therefore, various surrogate models mentioned above are summarized in Table 3, which shows that Kriging model, NN model, and RBF model are the most commonly-used ones. However, there is no more systematic comparative study to explore the performance of these different surrogate models.

\begin{table}[h]
\small
\begin{center}
\caption{Surrogate models used in the hull form optimization.}
\begin{tabular}{cc} \hline
Surrogate model & Articles \\  \hline
RSM & \cite{zong2018hull,lin2019automatic,guan2021parametric,wan2022interval,li2014bow}\\
GP-RSM & \cite{coppede2019hydrodynamic} \\
GP & \cite{rafiee2022multi,wang2023study,renaud2022multi,pellegrini2023multi} \\
Kriging & \cite{coppede2019hydrodynamic,wu2017neumann,miao2020cfd,liu2022resistance,miao2020hull,liu2017multiple,lin2018hull,tahara2008computational,zhang2021kriging} \\
Elman NN & \cite{zhang2018computational} \\
ANN (BP) & \cite{hou2017minimum,ccelik2021reduced,wu2021uncertain,hou2017hull,chen2021multi,diez2015multi} \\
RBF & \cite{serani2016ship,huang2016hull,liu2021hull,wang2021many,chen2015high,qiang2022optimization,huang2016new,wang2021aerodynamic,wei2019hull,tian2021multi,leotardi2016variable,chi2016overview} \\
DRBF  & \cite{luo2021hull,diez2018stochastic} \\
SRBF  &  \cite{serani2022hull,volpi2015development,pellegrini2022multi,piazzola2023comparing}\\
SVR  & \cite{zhu2019optimized,feng2018multidisciplinary}\\
SVM-GSM  &  \cite{mao2022multi}\\
Co-Kriging  & \cite{liu2023multi,liu2022multi,gaggero2022marine} \\
Surrogate integration method  & \cite{hu2022conceptual}  \\
Deep Belief Networks  & \cite{zhang2022research,zhang2021research}\\
\hline
\end{tabular}
\end{center}
\end{table}

\subsection{Application cases}

\subsubsection{Total drag optimization case of DTMB-5415 hull}

DTMB-5415 (David Taylor Model Basin) is a preliminary design of a warship by the U.S. Taylor Pool (DTMB) in the 1980s, with a sonar cover on the bow and a square transom on the stern, and has become an internationally recognized standard ship model with rich test results for numerical calculation verification. The ship belongs to medium and high speed ship. Its three-dimensional model is shown in Figure 4-7.

\begin{figure}[ht]
\begin{center}
    \subfigure[main view]{\includegraphics[width=0.45\textwidth]{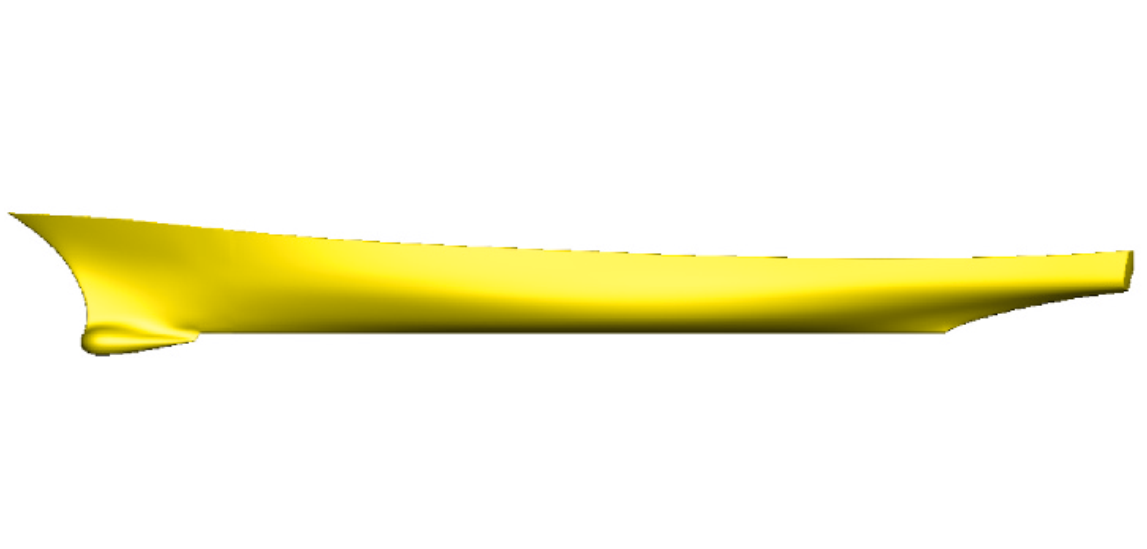}}
    \quad
    \subfigure[global view]{\includegraphics[width=0.4\textwidth]{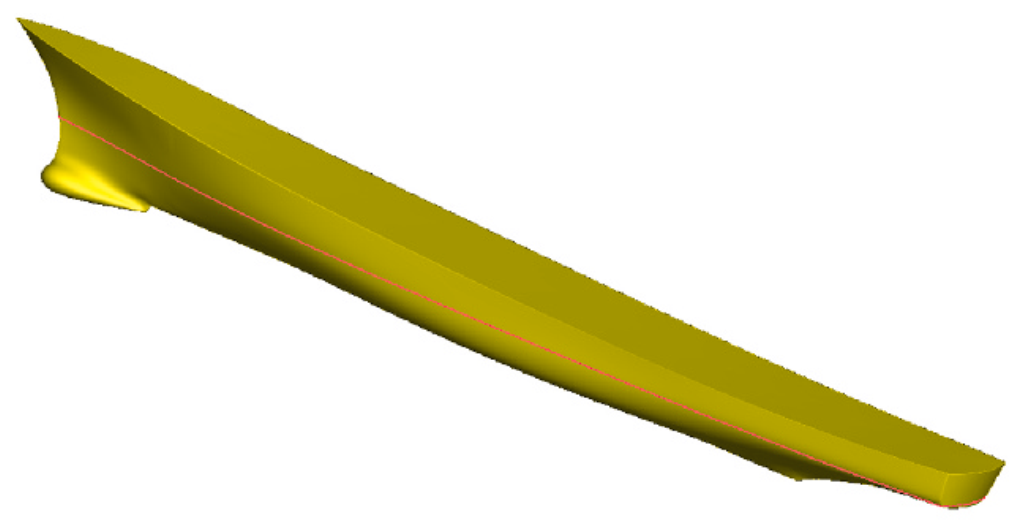}}
    \caption{Geometry model of DTMB-5415.}
\end{center}
\end{figure}

\begin{table}[h]
\begin{center}
\caption{Summary of the optimal results in terms of the objective function.}
\begin{tabular}{cccc} \hline
/ & Model prediction (N) & CFD calculation (N) & Reduction ratio \\  \hline
Initial & / & 43.220 & / \\
Opt-K   & 39.500 & 42.011 & 2.80\% \\
Opt-CoK & 41.020 & 41.001 & 5.13\% \\
\hline
\end{tabular}
\end{center}
\end{table}

\subsubsection{Total drag optimization and Wake fraction reduction of JBC ship}

\begin{figure}[ht]
\begin{center}
    \subfigure[main view]{\includegraphics[width=0.45\textwidth]{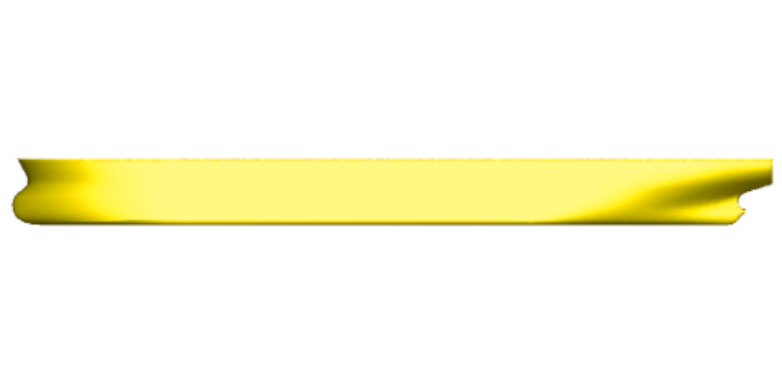}}
    \quad
    \subfigure[global view]{\includegraphics[width=0.4\textwidth]{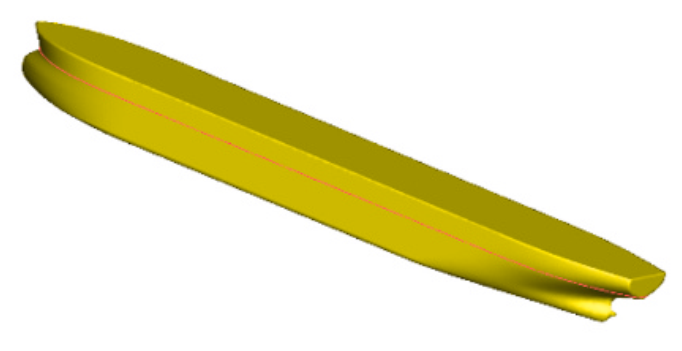}}
    \caption{Geometry model of DTMB-5415.}
\end{center}
\end{figure}

According to the Kriging models 1 and 2 and the Co-Kriging models 3
and 4 discussed in the preceding section, the second-generation Nondominated Sorting Genetic Algorithm (NSGA-II) is used to solve the multi-objective hydrodynamic performance optimization problem. 

Based on the constructed Kriging and Co-Kriging surrogate models, the optimal Pareto solution sets of the optimization problem can be obtained using the NSGA-II, and each point on the Pareto front (corresponding to each new hull form) has different degrees of performance improvement (objective function value decrease) compared with the initial hull.

\begin{table}[h]
\begin{center}
\caption{Summary of the optimal results.}
\begin{tabular}{ccccccc} \hline
 & \textbf{Model}  & \textbf{CFD}  & \textbf{Reduction}  & \textbf{Model}  & \textbf{CFD}  & \textbf{Reduction}  \\  
 & prediction (N) & calculation (N) & ratio & prediction (N) & calculation (N) & ratio \\ \hline
Initial & / & 32.702 & / & / & 0.347 & / \\
Opt1-K & 30.361 & 31.206 & 4.57\% & 0.325 & 0.328 & 5.48\% \\
Opt2-K & 30.787 & 30.532 & 6.64\% & 0.308 & 0.322 & 7.20\% \\
Opt3-CoK & 29.402 & 30.847 & 5.67\% & 0.300 & 0.319 & 8.07\% \\
Opt4-CoK & 29.806 & 31.021 & 5.14\% & 0.297 & 0.311 & 10.37\% \\
\hline
\end{tabular}
\end{center}
\end{table}

\section{Multidisciplinary design optimization and robust design of ship hull}

Ship form optimization design based on hydrodynamic theory is a systematic discipline, which needs the support of many disciplines. During the optimization process, it is necessary to consider the coupling of multiple disciplines, the design of variables and various nonlinear constraints. As a result, Multidisciplinary Design Optimization (MDO) \cite{martins2013multidisciplinary} strategy has been successfully applied in the field of ship hull form design and continues to gain attention. MDO was first proposed in the aerospace field and applied to the exterior design of aircraft (e.g., airplane wings). Then, MDO was expanded to various complex engineering system optimization problems. For example, MDO used for hull form optimization takes full account of the interaction and coupling among various disciplines/subsystems-such as ship hull form (hull scheme and hydrodynamic performance analysis), structural science (pressure-resistant hull design) and energy science (energy consumption and energy carrying capacity). More specifically, MDO utilizes effective design optimization strategies to optimize all design variables synchronously and coordinate interdisciplinary interactions to organize and plan the entire design process of the hull form. Usually, MDO can be mainly categorized into single-level and multi-levels according to the multidisciplinary system decomposition hierarchy. Single-level algorithms are suitable for solving simple systems with few design variables, small computational volume and less disciplines. Whereas multi-level algorithms can make up for the above mentioned deficiencies effectively. Commonly-used multi-level optimization algorithms \cite{martins2013multidisciplinary} include Collaborative Optimization (CO), Concurrent Subspace Optimization (CSO), Bilevel Integrated System Synthesis (BLISS) and Enhanced Collaborative Optimization (ECO) algorithms. The development of these algorithms intends to be mature, and readers interested in can refer to \cite{martins2013multidisciplinary} for more details.

For MDO, design variables and parameters are usually regarded as deterministic input information. However, uncertainties are widespread and unavoidable exist in real sea conditions, such as aleatoric uncertainties caused by random factors and epistemic uncertainties caused by lack of knowledge \cite{esmailian2022ship,li2023ship}. For example, material properties, geometry and external workloads of equipment are all uncertain. For multidisciplinary system, uncertain information can be propagated through the coupling relationship between different disciplines. The cumulative effects of the propagation can degrade the performance of the system and thus greatly reduce the reliability and safety of the system. Combined with uncertainty analysis in the field of engineering optimization, Figure 5 gives three common types of uncertainty faced in robust design optimization of ships.

\begin{figure}[h]
\begin{center}
\includegraphics[scale=0.95]{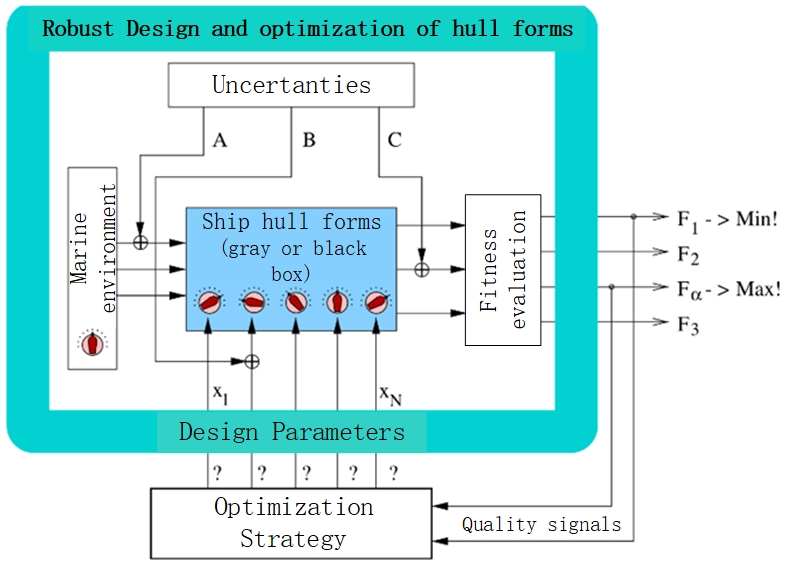}
\caption{Diagram of uncertainty types in robust ship design systems.}
\label{Fig5}
\end{center}
\end{figure}

\begin{enumerate}
\item Environmental uncertainty (A in Fig. 5). It is possible that the same vessel performs differently in different sea state environments (e.g., the effects of wind, waves, and currents). The actual speed varies from the designed speed to some extent.
\item Parameter space uncertainty (B in Fig. 5). This kind of uncertainty exist in decision space. The performance of optimal solution may degrade due to the error of the manufacturing techniques. The goal of robust optimization is to find the solution with little performance loss within the error range of the decision variable $\textbf{x}$.
\item Performance evaluation uncertainty (C in Fig. 5). It refers to uncertainty in the target space, caused by simulation or experimental results with noise or bias in the predictions of the approximate model \cite{wan2022interval}. Robust optimization aims to deal with this kind of uncertainty.
\end{enumerate}

Some scholars have applied MDO to scientific ship hull form design. For example, Liu \textit{et al}. \cite{liu2017multiple} established a multi-objective MDO design for heavy underwater vehicles. Feng \textit{et al}. \cite{feng2018multidisciplinary} considered the multidisciplinary optimization of ship hull forms for offshore aquatic vessels, in which both the drag performance and the quality of the aft flow field were taken into account. Chen \textit{et al}. \cite{chen2021multi} combined the coupling of hydrodynamic discipline, weight and balance discipline, propulsion system discipline, and energy discipline, thus they established a four-disciplinary, three-objective optimization system of drag, forward momentum, and endurance. Luo \textit{et al}. \cite{luo2021hull} established a MDO system based on CO method for underwater vessels. Hu \textit{et al}. \cite{hu2022conceptual} combined the disciplines of hull shape, structural design, and energy utilization to establish a multidisciplinary coupled computational model and conducted optimization by PSO. Yang \textit{et al}. \cite{yang2022new} proposed a MDO framework based on CCO method, which considers the underwater glider optimization in five disciplinary areas of hydrodynamic shape, pressure-resistant shell, buoyancy, attitude and energy.

As seen from Figure 5, ship robust design are fronted with many uncertainties. Measuring Uncertainty Quantification (UQ) is a prerequisite for achieving robust design optimization, which can address the effects of on output brought by uncertain inputs. Some commonly-used UQ methods include multidimensional polynomial chaos \cite{wei2019hull}, interval analysis \cite{wan2022interval}, and traditional Monte Carlo method. For example, Wei \textit{et al}. \cite{wei2019hull} established reliability-based robust design optimization (RBRDO) for ship hull form design and performed uncertainty analysis by combining polynomial chaos expansion and maximum entropy method; Wan \textit{et al}. \cite{wan2022interval} used interval analysis for UQ analysis and a two-tier nested optimization system where the inner tier considers the uncertainty interval of the objective function. In addition, they considered the uncertainty of the surrogate models and performed UQ based on the confidence of the surrogate models. Other related works that consider UQ in the field of ship design can be found in detail in the literature \cite{serani2022hull,wang2023study,diez2018stochastic,volpi2015development,leotardi2016variable,piazzola2023comparing}.

In terms of MDO for ship design, Diez \textit{et al}. \cite{diez2010multidisciplinary} analyzed some uncertain factors and applied Bayesian theory to quantify the degree of uncertainty of optimal design, and then established the conventional model and the model based on robust design in speedboat design. Leotardi \textit{et al}. \cite{leotardi2016variable} considered multidisciplinary robust design optimization (MRDO) in uncertain environments. They achieved optimal ship hull form design assisted by variable accuracy approximation models through a multidisciplinary feasibility structure. Wu \textit{et al}. \cite{wu2021uncertain} considered the design optimization of ultra-deep sea ship design system, where MDO was divided into two parts, i.e., deterministic optimization and uncertainty optimization. They used the deterministic CO structure to organize the coupled collaborative optimization of related disciplines, and used the interval analysis method to introduce the uncertainty of design parameters and surrogate model. Finally, more reliable optimization results were obtained through PSO optimization.

In conclusion, although MDO has achieved some success in the field of ship hull form design, there are still a lot of aspects that need to be further studied. How to combine intelligent optimization algorithms and approximation techniques to achieve faster modeling, higher optimization efficiency, and better optimization quality. Moreover, robust MDO method incorporating uncertainty can fully consider the actual large engineering system, which is conducive to improving the practicality and reliability of the ship hull form. It is promising to provide guidance for exploring the new generation of ship hull form design methods.

\section{Advances in surrogate-assisted evolutionary optimization methods for hull design}

Evolutionary optimization methods are being increasingly studied for solving hull form design optimization for its easy implementation and high efficiency. However, the objective function evaluations of hull design optimization are usually time-consuming (i.e., computationally expensive). The representative methods used are surrogate-assisted evolutionary algorithms (SAEAs) \cite{guo2018heterogeneous,cui2021surrogate}, also referred to data-driven evolutionary optimization \cite{jin2018data}. The basic framework of SAEAs is illustrated in Fig.6(a), where the surrogate model is trained by historical or real-time data of the optimization problems, and they can be used to replace a majority part of actual models for the purpose of the rapid fitness evaluation. Nowadays, SAEAs have been the mainstream method in the field of ship form design optimization, which are expected to provide a guidance for future work of artificial intelligence-based and knowledge-based ship hull form optimization.

\begin{figure}[ht]
    \centering
    \subfigure[General framework of SAEAs]{\includegraphics[width=0.4\textwidth]{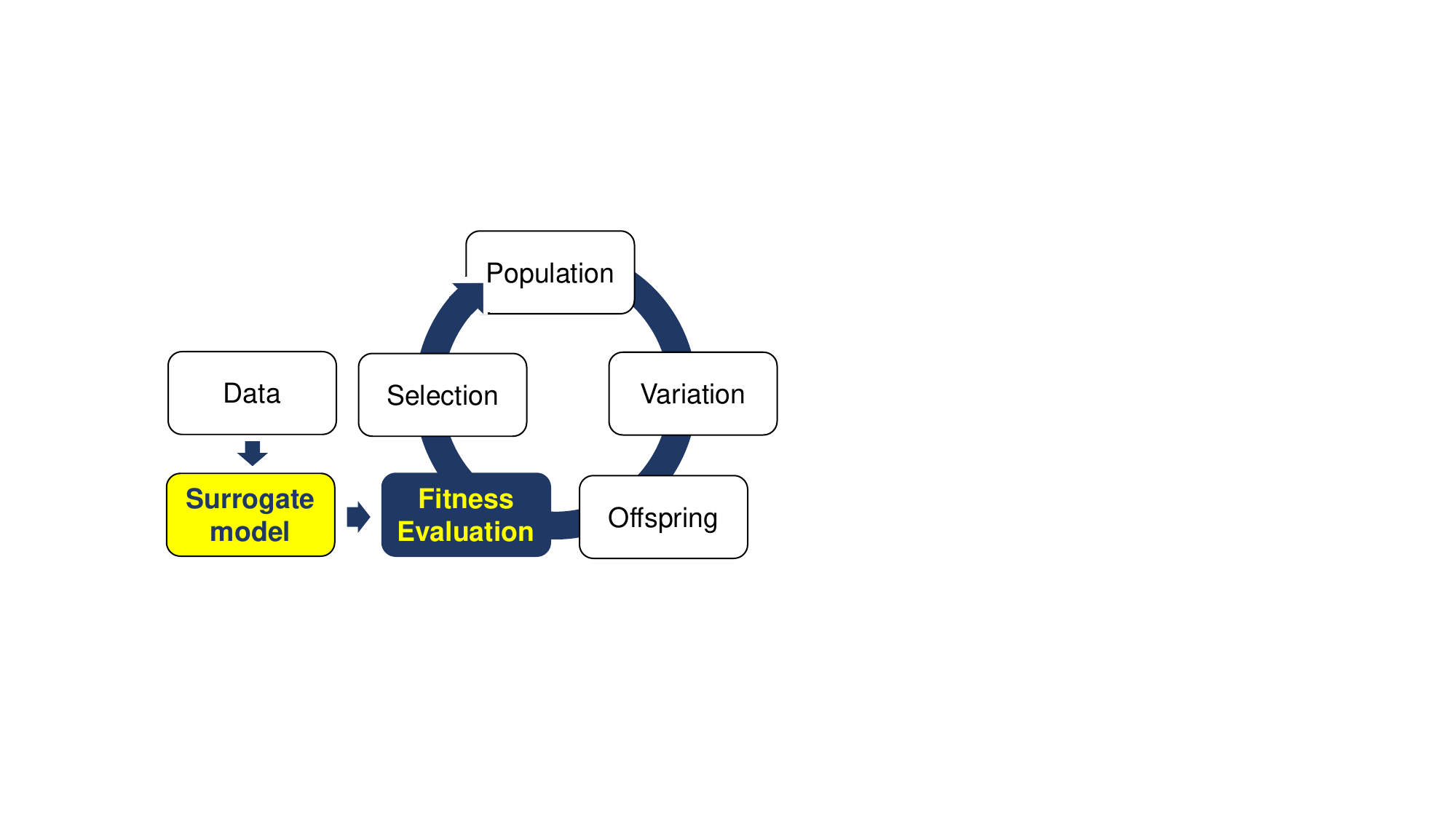}}
    \quad
    \subfigure[Surrogate model construction]{\includegraphics[width=0.4\textwidth]{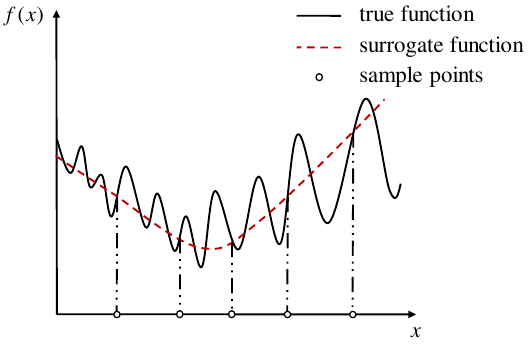}}
    \caption{The illustration of SAEAs.}
\end{figure}

The first step of SAEAs is surrogate model construction, as illustrated in Fig.6(b), for which Gaussian process (also known as Kriging model) \cite{liu2020gaussian} is widely used \cite{guo2018heterogeneous} as a result of its good performance. GP model can provide a confidence-based uncertainty measurement, which facilitates the online model updating. However, existing studies have shown that the time complexity of constructing a GP grows exponentially and the accuracy of the model decreases as the dimension of the decision variables increases, making it difficult to solve large-scale optimization problems \cite{cui2021surrogate,cui2022bi}. In addition, the fitting ability of different models trained by data with different characteristics varies greatly. It is impossible to construct a unified and efficient surrogate model. Recently, SAEAs for expensive multi-objective optimization problems have gradually gained attention from scholars. A number of works have appeared both at the level of algorithm design \cite{guo2018heterogeneous,cui2021surrogate} and engineering applications (e.g., trauma system allocation optimization \cite{wang2018random} and fuel efficiency \cite{bhattacharjee2019data}.).

Due to the deviation between surrogate model and real mode, it is necessary to update surrogate model by using the newly generated real data samples during the optimization process, which is called model management. In this procedure, surrogate model is updated by sampling new points through the extraction function according to the posterior information of the model. The purpose of the extraction function is to establish new sampling samples to update surrogate model, which plays a crucial role in balancing the optimization performance of both exploration and exploitation. Currently, model management in multi-objective optimization has become one of the frontiers, such as filling sampling criterion based on ensemble strategy \cite{guo2018heterogeneous}.

As a dominant method for solving expensive problems, how to construct effective and efficient surrogate models in high-dimensional space remains an immense challenge. A high-accuracy surrogate model can be trained by sufficient data samples, however, which can inevitably result in high surrogate construction time consumption. How to balance surrogate accuracy and construction time is a key issue when adopting SAEAs for solving HEPs. The existing surrogate models are mainly selected from GP, RF and RBF. Some novel and effective machine learning methods, such as deep brief networks \cite{yuan2022novel}, should be investigated as surrogate models in the evolutionary community, especially for problems with complicated data structures.

Recently, Jin \textit{et al}. \cite{jin2018data} conducted a detailed review and analysis of data-driven evolutionary algorithms. SAEAs are classified into online and offline methods according to whether new data are generated or not during the optimization process. Among which, online methods are more widely used as a result of promising optimization results. Some cutting-edge research on online SAEAs include \cite{jin2018data}: (1) improving model accuracy through surrogate model ensemble, (2) enhancing convergence by a combination of local and global surrogate models, (3) taking advantages of advanced machine learning techniques, such as semi-supervised learning, active learning and transfer learning during the whole process of evolutionary search. Therefore, investigation of these methods mentioned above is of promising potential for developing advanced SAEAs of green hull design.

\section{Conclusion and future directions}

This paper analyzes and summarizes various techniques of energy-efficient ship hull form optimization. Also, ship hull form design optimization based on SBD technology and intelligent optimization methods are presented, especially focusing on the latest development of intelligent optimization methods and surrogate model design. Meanwhile, the current development trend of multidisciplinary design optimization and data-driven evolutionary optimization theory research are analyzed to provide more guidance for its development in ship hull form design optimization research. Based on the above analysis of the developmental status of related research, the following points need to be further considered and studied in depth:
\begin{itemize}
    \item Deep synergistic optimization of hull form design with other ship performances is of great potential. The coupled synergy of multidisciplinary can be strengthened by multi-objective optimization, i.e., the establishment of a hull form with low energy consumption, low impedance, good endurance, and high robustness.
    \item To further explore intelligent optimization algorithms based on multi-fidelity surrogate models for hull form design. Multi-fidelity surrogate models can guarantee both high efficiency and high accuracy, but they have not been widely used in the optimization of hydrodynamic performance of ship hull forms.
    \item The uncertainties of numerical computation, real complex maritime conditions and ship operations need to be considered into optimization. Multidisciplinary domain knowledge of hull design can be integrated with machine learning algorithms driven by collected sample data. We should conduct in-depth research on constructing a hybrid knowledge- and data-driven optimization method for more energy-efficient hull form design.
\end{itemize}

\bibliographystyle{elsarticle-num-names} 
\bibliography{mybib}

\end{document}